\documentclass[journal=nalefd,manuscript=article,layout=twocolumn]{achemso}

\usepackage{chemformula} 
\usepackage[T1]{fontenc} 
\usepackage{sectsty}
\usepackage{float}
\sectionfont{\fontsize{12}{15}\selectfont}

\let\oldmaketitle\maketitle
\let\maketitle\relax
\author{Andrew Reynoso}
\affiliation[UC Berkeley]
{Department of Physics, University of California, Berkeley, 94720}
\author{Bohan Xu}
\affiliation[Penn State]
{Department of Physics, Pennsylvania State University, University Park, PA 16801}
\author{Vincent H. Crespi}
\email{vhc2@psu.edu}
\affiliation[Penn State]
{Departments of Physics, Chemistry, and Materials Science and Engineering, Pennsylvania State University, University Park, PA 16801}

\title[An \textsf{achemso} demo]
{Controlling nanothread backbone structure through precursor design}


\begin{document}

\begin{tocentry}\vspace{-4mm}\includegraphics{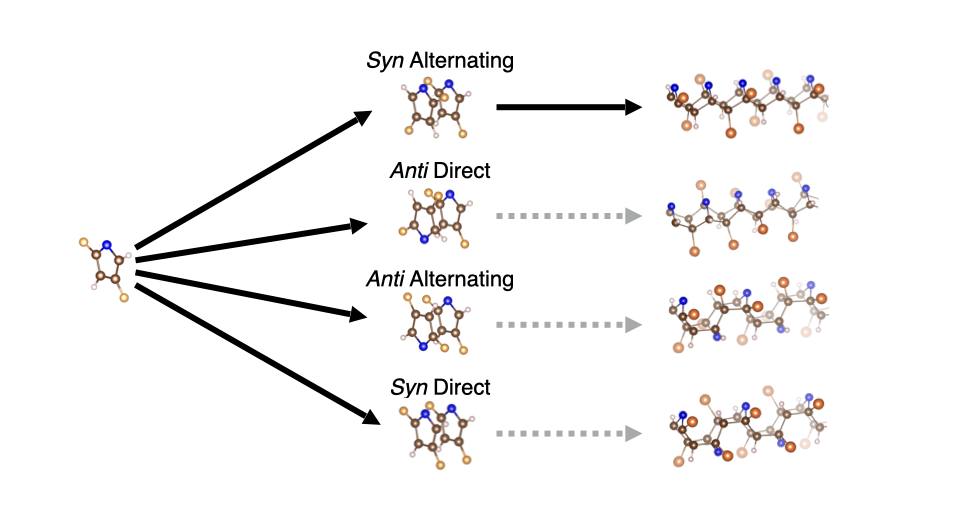}\end{tocentry}



\twocolumn[
\begin{@twocolumnfalse}
\oldmaketitle
\begin{abstract}
Nanothreads are 1D carbon-based nanomaterials produced by pressure-induced polymerization of multiply unsaturated (and typically aromatic) precursors with multiple bonds between adjacent precursors. We computationally design non-covalent interactions between functional groups on thread-forming monomers to control the relative stabilities of different nanothread backbones. In particular, functionalized furan or thiophene precursors are identified that favor nanothreads with oxygen or sulfur atoms arrayed along the same side of the thread backbone, rather than on alternating sides as currently seen experimentally for threads formed from unfunctionalized furan or thiophene. This heteroatom chain provides opportunities for unusual properties arising from a sterically compressed one-dimensional chain of p orbitals.
\end{abstract}
\end{@twocolumnfalse}
]


Carbon nanothreads are one-dimensional sp$^3$-bonded nanomaterials that form when precursors such as aromatic hydrocarbons polymerize along stacks at high pressure \cite{BenzeneExp2015, BenzeneExp2017, BenzeneTheory2001, Naph2019, FuncBenz2017}. Two such molecules are furan (C$_4$H$_4$O) and thiophene (C$_4$H$_4$S) \cite{FuranExp2021, ThioExp2019}. Growth through e.g. [4+2] cycloaddition of furan or thiophene monomers onto the end of a thread may align successive heteroatoms onto same or opposite sides of the backbone, configurations called \textit{syn} and \textit{anti}.  In the \textit{syn} case, the fully occupied (i.e. non-bonding) p orbitals of O or S overlap head-to-foot in a 1-D chain along the thread axis with a very short 2.6 -- 2.8 \AA\  interatomic separation, held in a state of compression that is balanced by a countervailing extensional stress in the torsionally and extensionally rigid sp$^3$ carbon backbone of the thread \cite{FuranThioTheory2022}. The electronic band that derives from this orbital stack significantly lowers the bandgap of saturated furan or thiophene-derived threads, could be electronically activated through e.g. p-type doping or optical excitation \cite{FuranThioTheory2022}, and couples to torsional or bending deformations that partially relax the steric compression. Unfortunately, the same steric repulsion that makes this stack so interesting also destabilizes the \textit{syn} configuration relative to its \textit{anti} competitors: \textit{anti} threads are the experimental product from compression of furan \cite{FuranNmr2021} and likely also thiophene \cite{ThioExp2019}. In an attempt to ``rescue'' the electronically and structurally interesting \textit{syn} configuration, we investigate computationally the relative energetics of various \textit{syn} and \textit{anti} backbone configurations where the precursors have been functionalized to express repulsive or attractive interactions that may favor \textit{syn} over \textit{anti}. These efforts help to build out our computational toolbox for nanothread backbone design by engineering non-covalent interactions into the thread precursors. 

\begin{figure}[h]
\includegraphics[width=0.95\columnwidth]{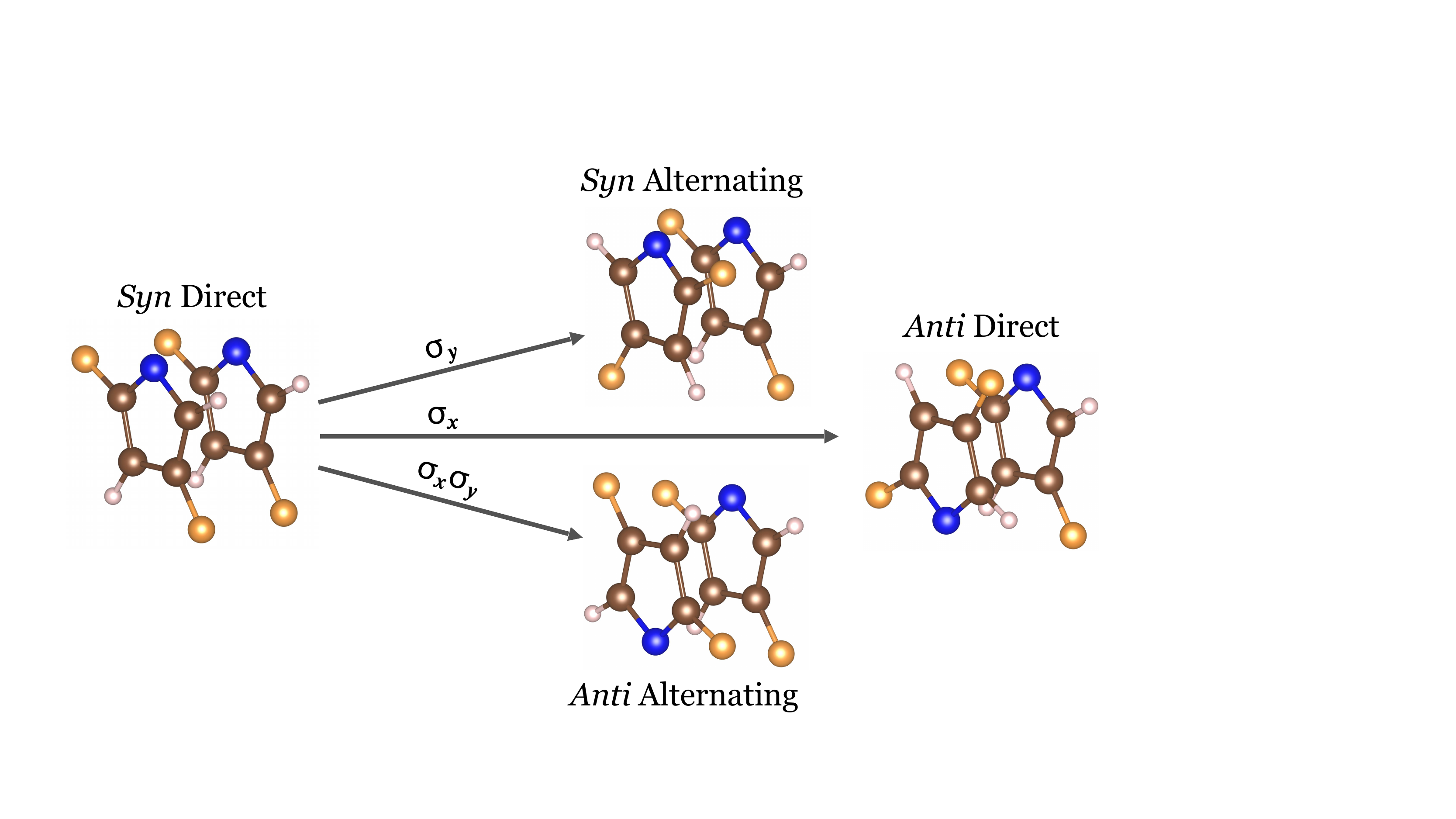}
\caption{Four distinct relative monomer orientations prior to [4+2] cycloaddition to form a nanothread. \textit{Syn}-direct connects monomers with identical orientation, while \textit{syn}-alternating, \textit{anti}-direct, and \textit{anti}-alternating reflect one monomer about $x$ (horizontal) and/or $y$ (vertical) axes as indicated, using (2,4)-dichlorofuran as an example.
\label{Fig-FourMonomerOrientations}}
\end{figure}


\begin{figure}[t]
\includegraphics[width=\columnwidth]{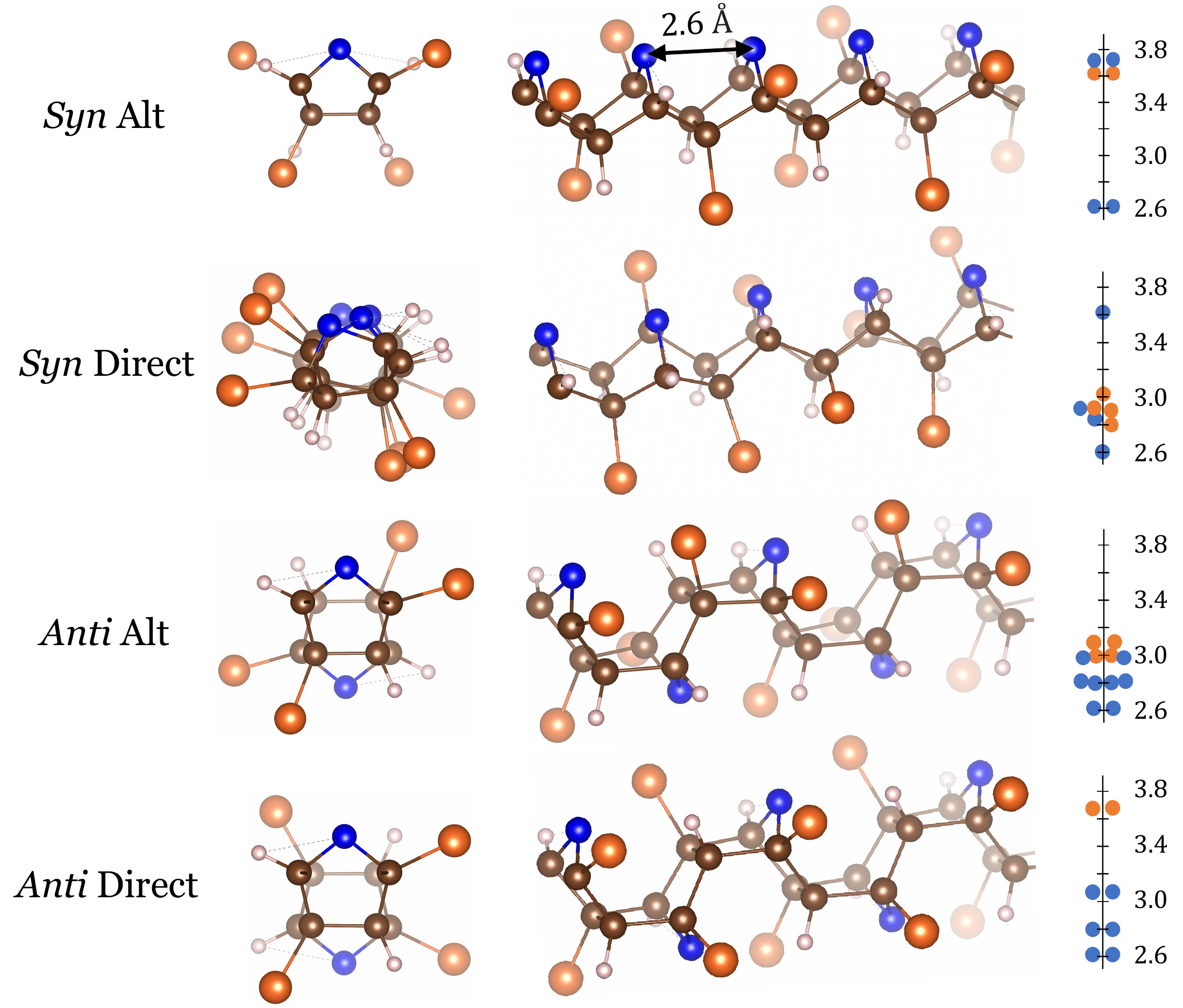}
\caption{Four threads resulting from the precursor alignments of Figure \ref{Fig-FourMonomerOrientations}. The small plots at right show chlorine–chlorine (blue) and chlorine–oxygen (orange) separations of less than 4 \AA. Here the \textit{syn}-alt thread better separates these atoms, reducing steric repulsion and contributing to the relative stability of that backbone geometry for a (2,4)-dichlorofuran precursor. 
\label{Fig-Four4+2ThreadTypes}}
\end{figure}

For each monomer considered, we examine nanothreads that may form by symmetry-allowed [4+2] cycloaddition \cite{FuranThioTheory2022}, as well as certain symmetry-disallowed [2+2] and [4+4] reactions.  We begin with an initial assessment within density functional tight binding (DFTB) of a combinatorially complete set of functional group attachment points (to a furan or thiophene core) and relative orientations of successive precursors along the thread axis, using the Amsterdam Density Functional package \cite{Adf, Adf2, AdfBasis} with DFTB3 parameter set  \cite{Dftb3a, Dftb3b, Dftb3c, Dftb3d}. This initial survey helps identify trends that may favor \textit{syn} backbones and focuses attention on promising thread geometries for further study within more precise first-principles density functional theory (DFT) using the Perdew–Burke-Ernzerhof (PBE) \cite{Pbe} exchange correlation functional with a plane-wave basis as implemented in the VASP software package \cite{VASPa, VASPb, VASPc, VASPd, VASPe}.

\begin{figure*}[h]
\includegraphics[width=1.3\columnwidth]{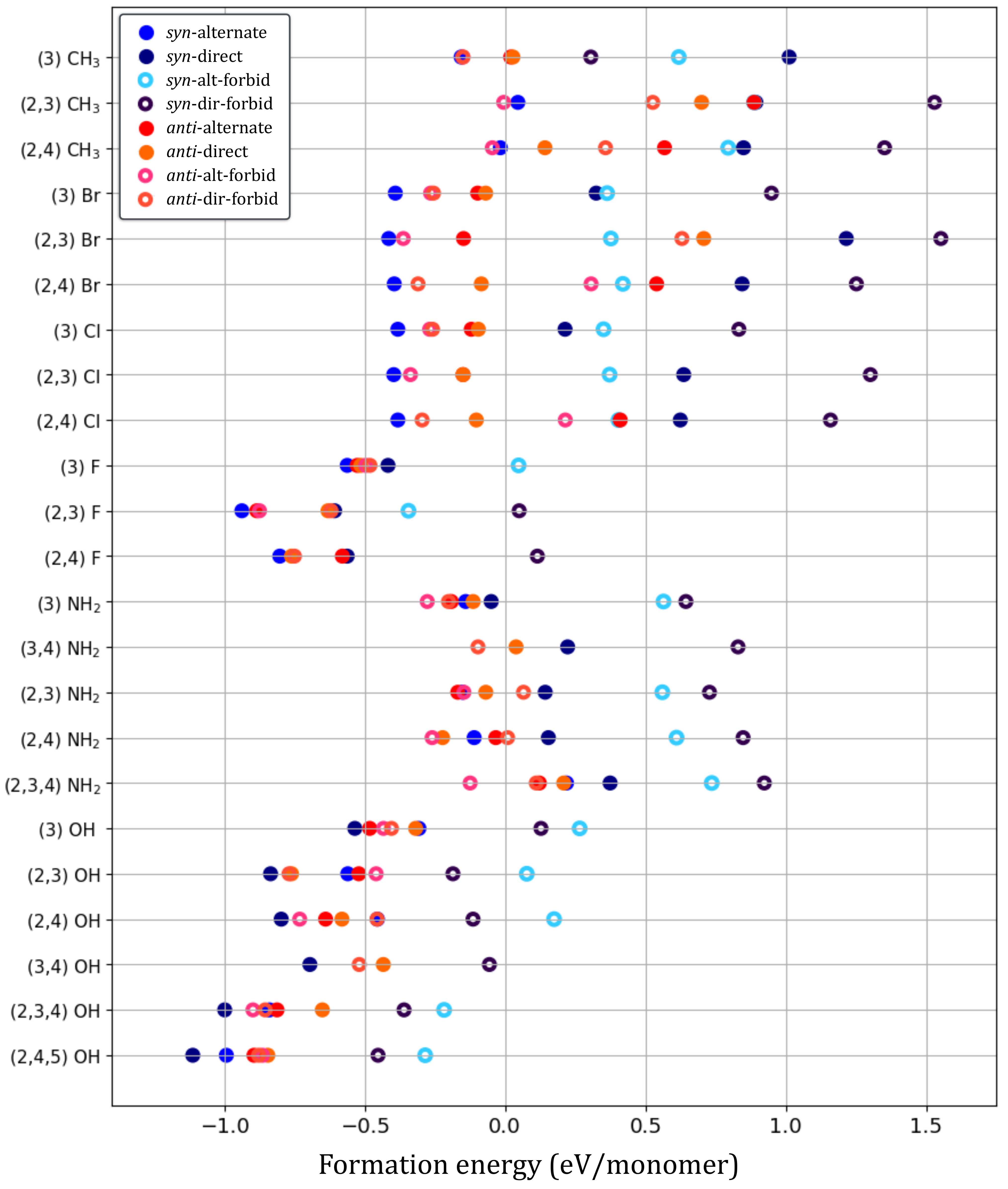}
\caption{Energies of furan-based nanothreads referenced to the isolated monomers within density functional theory using the PBE variant of the generalize gradient approximation. Sterically repulsive groups at (3), (2,3), and (2,4) tend to favor \textit{syn}-alternating threads.  Hydroxyl substitutions are broadly favorable for \textit{syn}-direct backbones, reflecting favorable hydrogen-bonded chains possible in this configuration. Backbones formed by [2+2]/[4+4] pathways (``-forbid'') rarely provide the most favorable structural isomer.
\label{Fig-DFTResults}}
\end{figure*}

For each functionalized precursor there are at most four distinct threads that can be produced by [4+2] cycloaddition if the threads are limited to a bond-topological periodicity of one or two monomers. Fixing the orientation of the first molecule without loss of generality, the possible  orientations for the next molecule are: identical to the first (\textit{syn} direct), a $\sigma_x$ reflection (\textit{syn} alternating), a $\sigma_y$ reflection (\textit{anti} direct), or $\sigma_x$ and $\sigma_y$ reflections (\textit{anti} alternating). These four possible precursor orientations, shown in Figure \ref{Fig-FourMonomerOrientations}, yield the four thread backbones of Figure \ref{Fig-Four4+2ThreadTypes}. (The ``direct'' and ``alternating''' distinctions are not needed for high-symmetry unfunctionalized furan and thiophene). For the most promising precursors, i.e. those that favor \textit{syn} threads under [4+2] cycloaddition, we further consider backbones formed by symmetry-forbidden [2+2] and [4+4] cycloadditions. Symmetry-allowed [4+4] cycloadditions are not considered in detail due to large backbone strains arising from the presence of heteroatoms within the ring. Threads formed by symmetry-allowed [2+2] plus symmetry forbidden [4+4] reactions are omitted on expectation of similar effects (and their combinatorial complexity). Note that the [2+2]/[4+4] threads that are considered in detail (the open circles in Figure \ref{Fig-DFTResults}) are almost always higher energy than others formed by [4+2] pathways; an illustrative example of this backbone structure is supplied in Supporting Information.


We focus first on furan-derived monomers. Given a side group (e.g. --Cl, --CH$_3$), preliminary DFTB computations suggested that substitutions at (3), (2,3), and (2,4) would tend to favor the \textit{syn}-alternating thread relative to the other three types, a result that is confirmed by the ab-initio results of Figure~\ref{Fig-DFTResults}. \textit{Syn}-direct threads with functional groups that express repulsive interactions are never energetically favored due to the short distance between them in adjacent monomers. When sites (3,4) or (2,5) are both occupied, the direct/alternating distinction is lost and \textit{syn}-alternating becomes as unfavorable as \textit{syn}-direct. However,  when positions (3) or (4) are occupied, one obtains steric repulsion between the oxygen lone pair and a halogen lone pair (or the methyl group) in all \textit{anti} threads, which disfavors them relative to \textit{syn}. Therefore, for functional groups with repulsive interactions the \textit{syn}-alternating threads become favorable when (3) or (4) is occupied but not both (and also not (2,5)). Only the (3), (2,3), and (2,4) configurations satisfy these conditions and thus favor \textit{syn}-alternating. The more favorable halogen--halogen and halogen--oxygen separations for \textit{syn}-alternating can also be seen in the plots on the right side of Figure \ref{Fig-Four4+2ThreadTypes}. Within these trends, larger substituents tend to generate more pronounced energetic distinctions. For example, 3-fluorofuran favors \textit{syn}-alternating by only 0.035 eV while 3-bromofuran favors \textit{syn}-alternating by 0.126 eV. Note that in the case of methyl, the \textit{anti}-alternating structure formed by the symmetry-forbidden [2+2]/[4+4] pathway is energetically competitive with the \textit{syn}-alternating thread formed by the symmetry-allowed [4+2] pathway, being slightly more favorable for (2,3) and (2,4) and slightly less favorable for substitution at (3).

%

\begin{figure}[t]
\includegraphics[width=0.75\columnwidth]{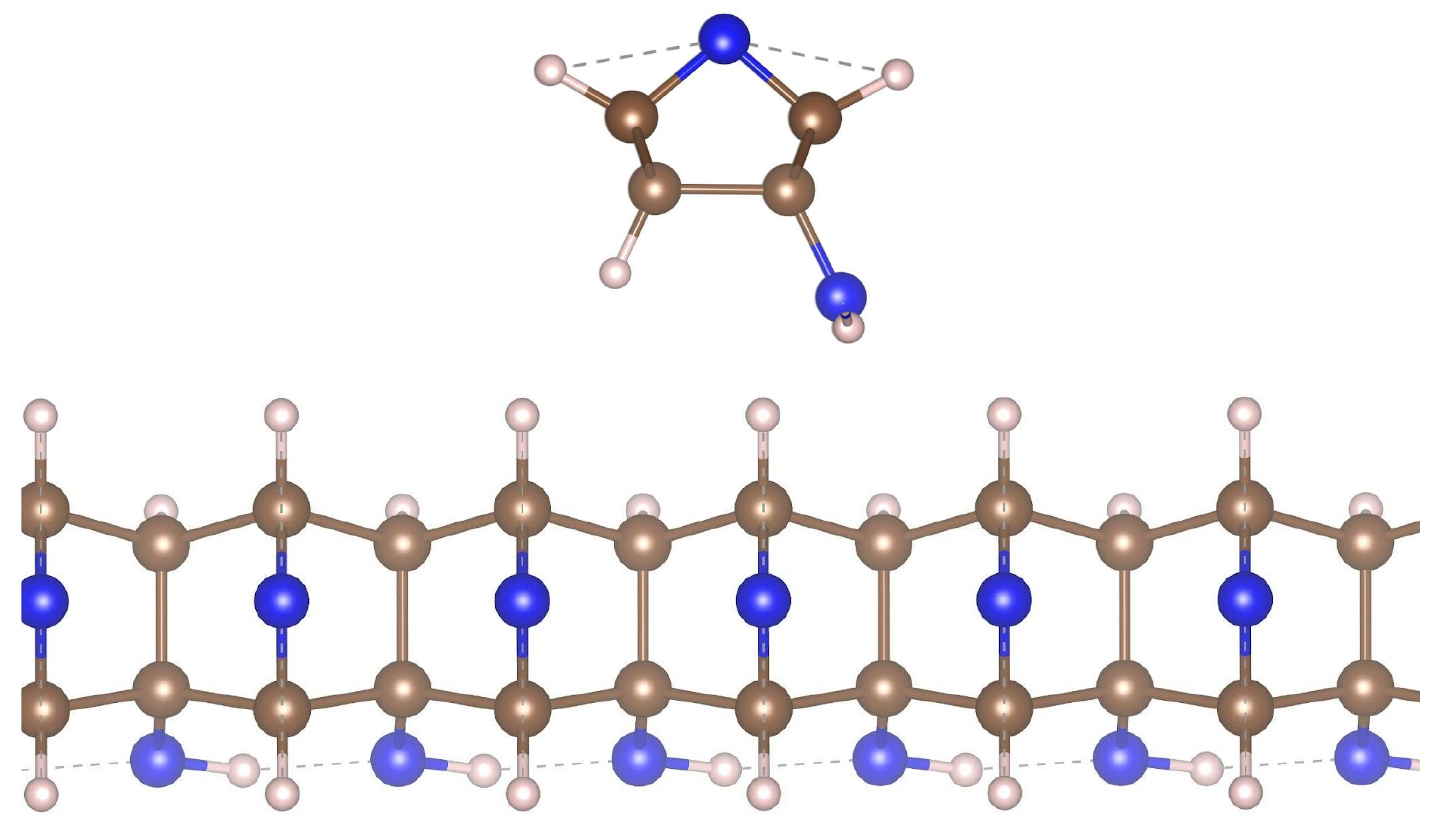}
\caption{The \textit{syn}-direct furan nanothread with a hydroxyl substitution at position (3), showing a favorable hydrogen bonding chain along the side of the thread that is not present for \textit{syn}-alt nor for either of the \textit{anti} configurations.
\label{Fig-OH}}
\end{figure}

Side groups capable of hydrogen bonding (--OH, --NH$_2$) may orient to express attractive interactions. Hydroxyl substitutions at (3), (2,3), (2,4), (3,4), (2,3,4), and (2,4,5) tend to favor \textit{syn}-direct, as this orientation aligns the functional groups favorably along the side of the thread as shown in Figure~\ref{Fig-OH}. Multiple hydroxyl substitutions tend to generate larger energetic distinctions between backbone geometries. Although --NH$_2$ was similarly favored energetically at the DFTB level, more precise first-principles calculations do not bear this out, as revealed by Figure~\ref{Fig-DFTResults} (i.e. no blue points at far left for those rows). This is likely because DFTB3 is known to describe the proton affinity of NH$_3$ inaccurately \cite{DFTB3NH3}. For completeness, the preliminary DFTB results are provided in Supporting Information, along with coordinate files for relaxed structures at the DFT/PBE level, including those formed by [2+2]/[4+4] pathways. 

Overall, multiple candidates for energetically favorable furan-derived \textit{syn} backbones are apparent based on precursor design. A complete theoretical design methodology would couple these efforts to molecular crystal structure prediction to identify favorable stacking geometries in precursor crystals and thus enable further studies of reaction kinetics in the high-pressure solid-state geometry; the current investigations set a foundation for the development of these further design capabilities. 

\begin{figure}[t]
\includegraphics[width=0.95\columnwidth]{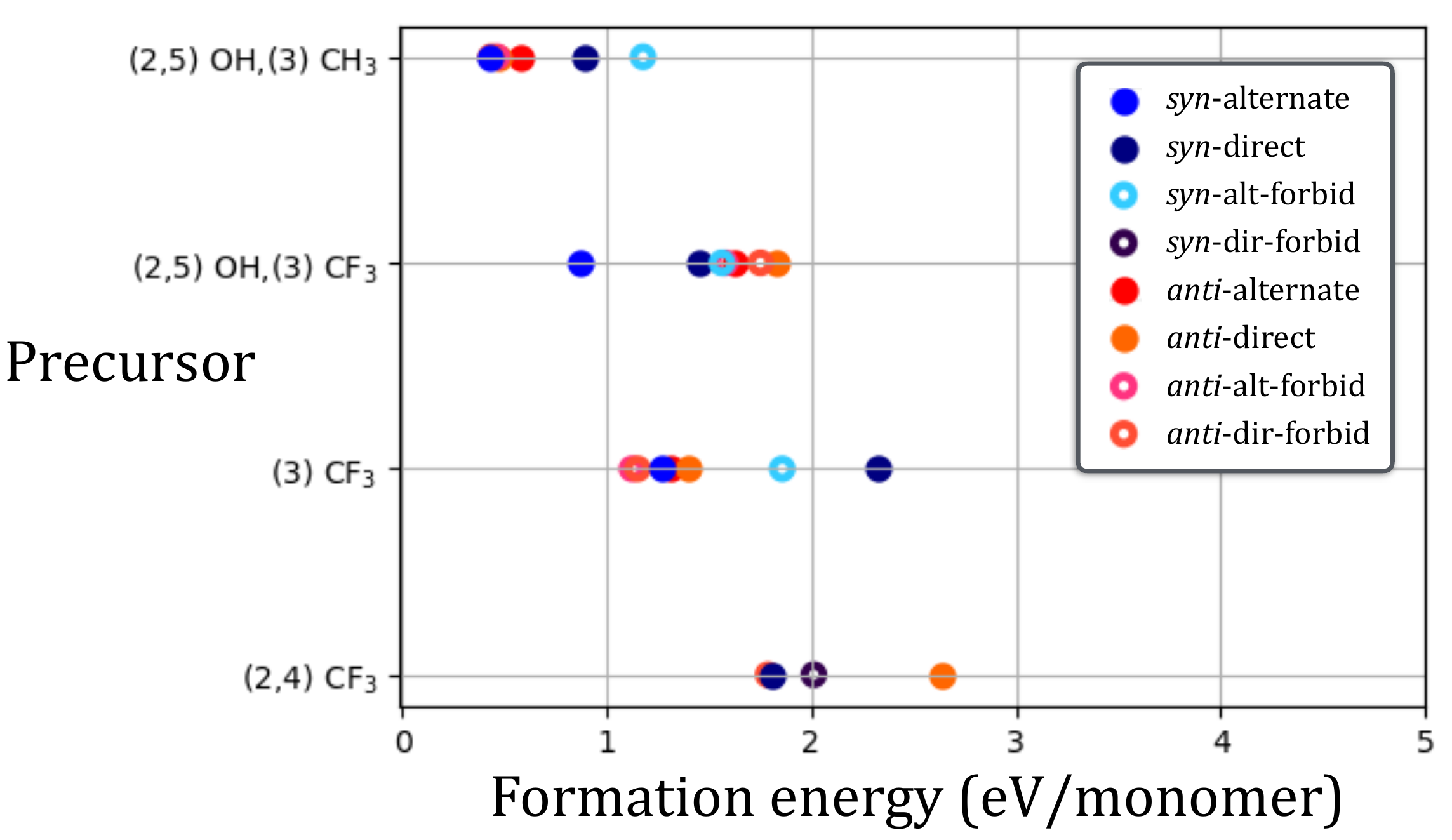}
\caption{Energies of functionalized thiophene-derived nanothreads that favor (or nearly favor) the \textit{syn} configuration.}
\label{Fig-ThiopheneDFTResults}
\end{figure}

\begin{figure}[t]
\includegraphics[width=0.85\columnwidth]{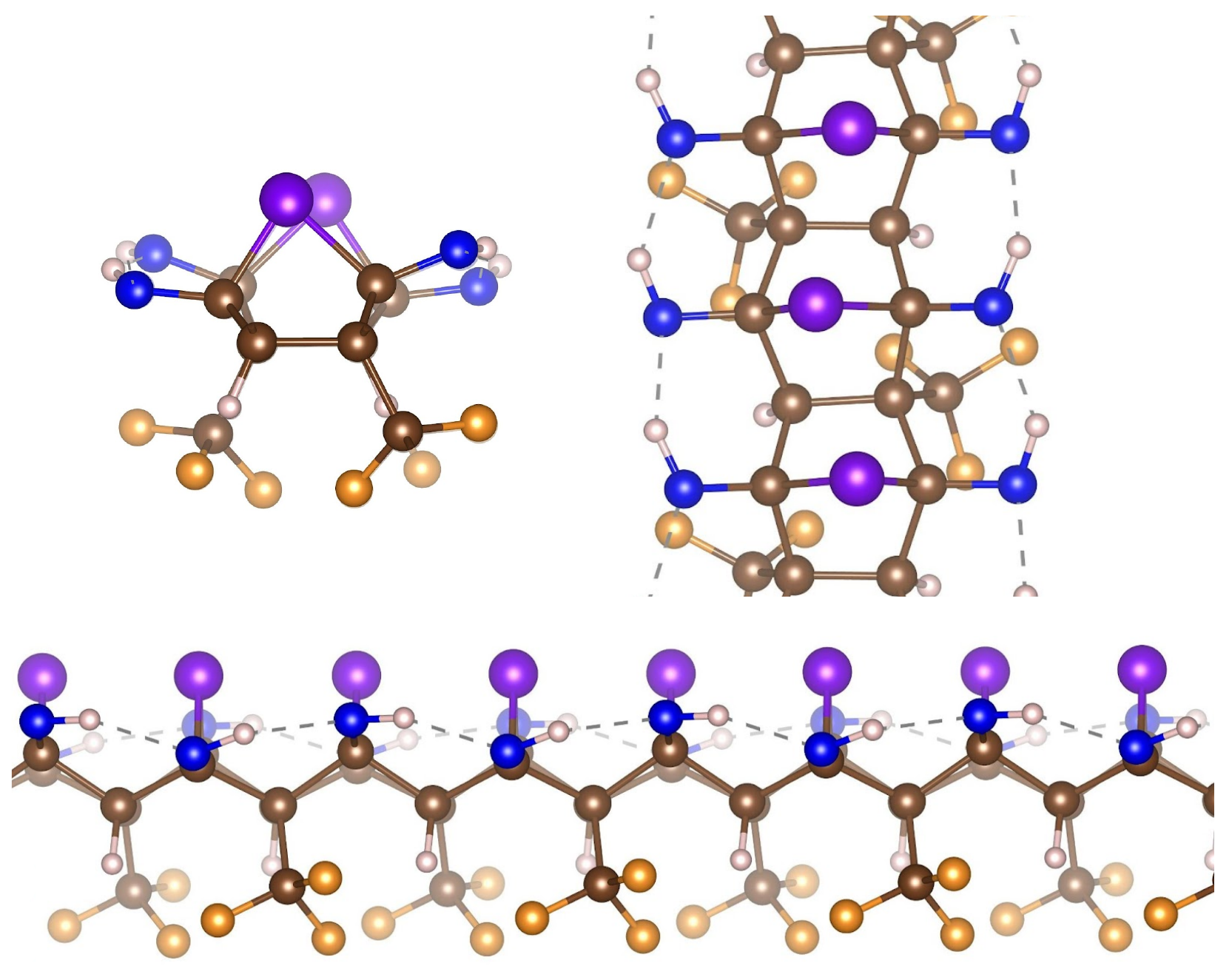}
\caption{A thiophene-derived monomer with --CF$_3$ substitution at (3) and --OH substitution at (2,5) strongly favors the \textit{syn}-alt configuration.
\label{Fig-ThiopheneCF3AndOH}}
\end{figure}

Thiophene presents a greater challenge in precursor design for backbone control, one that we were skeptical could be met due to the $\sim$1~eV/monomer energetic penalty of \textit{syn} relative to \textit{anti} for unfunctionalized, straight thiophene-derived threads \cite{FuranThioTheory2022}.  Nevertheless, successful candidates were identified. These make use of either a particularly bulky substituent (--CF$_3$) or a combination of repulsive (--CH$_3$ or --CF$_3$) and attractive (--OH) interactions to favor \textit{syn} over \textit{anti}.  More specifically, --CF$_3$ substitution at the (3) position and --CH$_3$ or --CF$_3$ substitution at (3) combined with hydroxyl at (2) and (5) all favor the \textit{syn} alternating backbone, as shown in Figures~\ref{Fig-ThiopheneDFTResults} and \ref{Fig-ThiopheneCF3AndOH}. The initial work with furan helped guide these investigations, particularly in selecting effective combinations of attractive and repulsive interactions.

\begin{figure}[t]
\includegraphics[width=0.8\columnwidth]{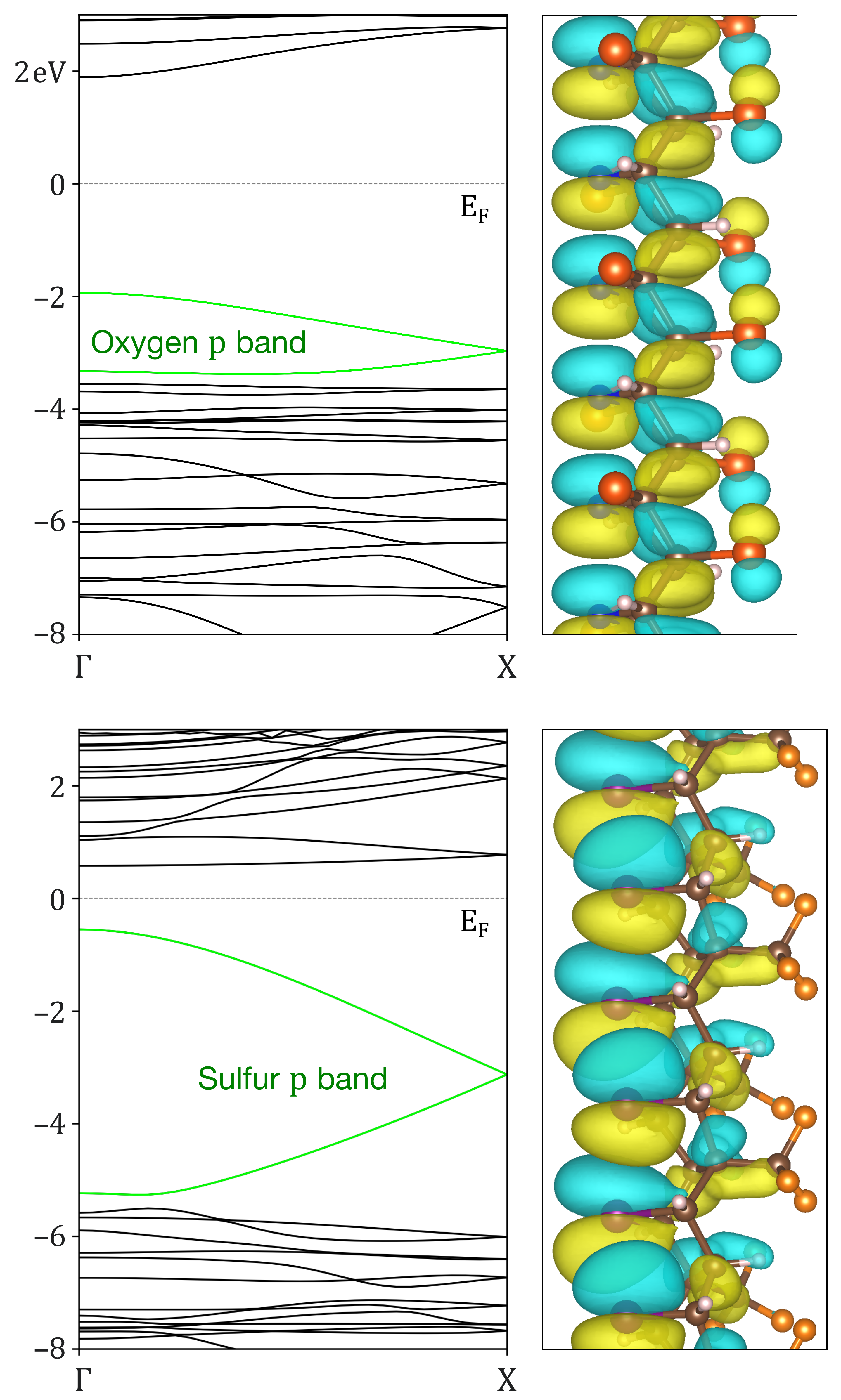}
\caption{Band structures of the 3-chlorofuran \textit{syn}-alternating and 3-CF$_3$ thiophene \textit{syn}-alternating nanothreads computed in density functional theory at the GGA PBE level. The bands deriving from the non-bonding heteroatom p orbital ``stack'' are highlighted in green and depicted  at right for the highest occupied state at $\Gamma$.
\label{Fig-Bands}}
\end{figure}

Given a nanothread with a \textit{syn} backbone, prospects for novel properties arise from the rigidity of the thread backbone enforcing the tight spacing of oxygen or sulfur heteroatoms. First, the system will have a net electric dipole moment orthogonal to the thread axis that will not easily self-compensate by dihedral twists or torsions of the thread backbone, potentially yielding a piezoelectric or ferroelectric nanothread crystal. Second, strong electronic overlap of O or S p orbitals oriented along the thread axis will produce a one-dimensional valence band of unusually high bandwidth as shown in Figure~\ref{Fig-Bands}, with a consequent reduction in the electronic bandgap. Threads with \textit{anti} backbones lack this highly dispersive state and thus have much larger gaps. Twists, bends, or other local deformations of the \textit{syn} backbone will couple strongly to this dispersive electronic state, opening prospects for strong electron-lattice coupling, polaronic conduction under hole doping, etc.

As noted earlier, these precursor design studies are a first step in the rational \textit{ab initio} design of desired nanothread backbone structures. The large number of viable candidates identified for attaining \textit{syn} furan-based threads and the ability to find candidates that even favor the highly compressed \textit{syn} thiophene backbone bode well for application of these principles, as they provide many candidates for further analysis of crystal packing and reaction kinetics in the solid state, noting that inter-thread interactions modulated by these functional groups will also play a role in overall reaction outcomes.


\begin{acknowledgement}

This work was funded by the Center for Nanothread Chemistry, a National Science Foundation (NSF) Center for Chemical Innovation (CHE-1832471).

\end{acknowledgement}

\begin{suppinfo}

The following files are available free of charge.
\begin{itemize}
  \item SI.docx: [2+2]/[4+4] backbone structures, computational methods, DFTB screening results
  \item AtomicCoordinates.pdf: Structural coordinates relaxed at the DFT/PBE level
\end{itemize}

\end{suppinfo} 
 
%
%

\bibliography{Bibliography}

\end{document}